\documentclass[10pt]{article}

\usepackage{amsmath}
\usepackage{amssymb}

\usepackage{graphicx}

\usepackage{cite}

\usepackage{color}

\usepackage[labelfont=bf,labelsep=period,justification=raggedright]{caption}

\newcommand{\ao}{\alpha_{1}}
\newcommand{\at}{\alpha_{2}}
\newcommand{\att}{\alpha_{3}}
\newcommand{\af}{\alpha_{4}}
\newcommand{\bo}{\beta_{1}}
\newcommand{\bt}{\beta_{2}}

\newcommand{\cs}{\left(1 - \frac{S}{K_{S}} \right)_{+}}

\makeatletter
\renewcommand{\@biblabel}[1]{\quad#1.}
\makeatother

\date{}

\begin{document}

% Title must be 150 characters or less
\begin{flushleft}
{\Large
\textbf{The Role of Osteocytes in Targeted Bone Remodeling: A Mathematical Model}
}
% Insert Author names, affiliations and corresponding author email.
\\
Jason M.\ Graham$^{1, \ast}$,
Bruce P.\ Ayati$^{2,3,5}$,
Sarah A.\ Holstein$^{4}$,
James A.\ Martin$^{5}$
\\
\bf{1} Department of Mathematics, University of Scranton, Scranton, PA, USA
\\
\bf{2} Department of Mathematics, University of Iowa, Iowa City, IA, USA
\\
\bf{3} Program in Applied Mathematical and Computational Sciences, University of Iowa, Iowa City, IA, USA
\\
\bf{4} Department of Internal Medicine, University of Iowa, Iowa City, IA, USA
\\
\bf{5} Department of Orthopaedics and Rehabilitation, University of Iowa, Iowa City, IA, USA
\\
$\ast$ E-mail: jason.graham@scranton.edu
\end{flushleft}

% Please keep the abstract between 250 and 300 words
\section*{Abstract}
Until recently many studies of bone remodeling at the cellular level have focused on the behavior of mature osteoblasts and osteoclasts, and their respective precursor cells, with the role of osteocytes and bone lining cells left largely unexplored. This is particularly true with respect to the mathematical modeling of bone remodeling. However, there is increasing evidence that osteocytes play important roles in the cycle of targeted bone remodeling, in serving as a significant source of RANKL to support osteoclastogenesis, and in secreting the bone formation inhibitor sclerostin. Moreover, there is also increasing interest in sclerostin, an osteocyte-secreted bone formation inhibitor, and its role in regulating local response to changes in the bone microenvironment.

Here we develop a cell population model of bone remodeling that includes the role of osteocytes, sclerostin, and allows for the possibility of RANKL expression by osteocyte cell populations. We have aimed to give a simple, yet still tractable, model that remains faithful to the underlying system based on the known literature. This model extends and complements many of the existing mathematical models for bone remodeling, but can be used to explore aspects of the process of bone remodeling that were previously beyond the scope of prior modeling work. Through numerical simulations we demonstrate that our model can be used to explore theoretically many of the qualitative features of the role of osteocytes in bone biology as presented in recent literature.
% Please keep the Author Summary between 150 and 200 words
% Use first person. PLoS ONE authors please skip this step.
% Author Summary not valid for PLoS ONE submissions.
%\section*{Author Summary}

\section*{Introduction}
The skeleton is a highly dynamic organ that undergoes constant remodeling throughout life. Remodeling is involved in systemic mineral homeostasis and is required for the routine repair of microfractures and for adaptation to mechanical stress. Even after the skeleton reaches its adult size, bone remains metabolically active. Bone remodeling involves the coupling of bone resorption and bone formation, and diseases are characterized by imbalances between formation and resorption. This includes such common disorders as osteopenia and osteoporosis \cite{ryser2010}, as well as malignant processes such as multiple myeloma and other cancers that metastasize to the bone \cite{hofbauer2004a,hofbauer2004b}. As these disorders contribute to morbidity and mortality in millions of people worldwide, there is great interest in improving our understanding the processes that regulate bone remodeling.

Bone remodeling is characterized by complex mechanical and biochemical signaling pathways. Remodeling typically occurs for one of three reasons: to maintain mineral homeostasis; to adapt to mechanical changes; and to repair damage \cite{burr2002,parfitt2002}. Remodeling related to mineral homeostasis need not occur at a specific site, and is therefore referred to as non-targeted remodeling \cite{burr2002,parfitt2002}. In contrast, remodeling due to mechanical changes or to damage is site-specific and is termed targeted remodeling \cite{burr2002,parfitt2002}. The key cellular players in targeted bone remodeling include the osteoblasts and osteoclasts. There is also increasing evidence to support the role of osteocytes as active participants in this remodeling process \cite{bonewald2011,neve2012}.

Osteocytes are mature osteoblasts that become embedded within the bone matrix during the formation phase of bone remodeling. During this process osteocytes develop cytoplasmic processes which run through the canaliculi, forming a communication network that can convert mechanical signals into biochemical signals \cite{bonewald2011,kular2012,neve2012}. Osteocytes are the most abundant and long-living bone cells and are found throughout the skeleton, in contrast to the much shorter lived osteoblasts and osteoclasts which are only transiently found on the surface of the bone. Osteocytes actively secrete growth factors that stimulate bone formation, as well as sclerostin, which inhibits bone formation \cite{heino2009,kular2012,neve2012}. The complex relationship between osteocytes, osteoblasts, osteoclasts, and their precursors, is continuing to be elucidated.

A number of mathematical and computational models of various aspects of bone physiology have been proposed. Many of these models focus on either bone fracture healing, \cite{Geris4,Geris2,Geris3,Geris5,Geris1,parfitt2002}, or bone remodeling, \cite{ayati2010,buenzli2012b,buenzli2012a,buenzli2011,defranoux2005,graham2012,komarova2005,komarova2003,lemaire2004,liotier2012,buckwright2004,buckwright2005,pivkom2010a,pivonka2008,pivonka2010b,ryser2010,ryser2009,zumsande2011}. These models take different approaches, and focus on various aspects of the remodeling process. The early modeling approach described in \cite{komarova2005,komarova2003} makes use of the biochemical systems analysis formalism of Savageau (\emph{e.g.}, \cite{savageau1,savageau2,savageau3,savageau1976,voit2000}) developed to represent bone remodeling as the effect of changes in bone cell populations; a so-called ``cell population'' model. These models are comprised of a system of two ordinary differential equations (ODEs) representing the cell population dynamics of osteoclasts and osteoblasts. In addition, they are chemically implicit in that they do not track chemical quantities as functions of time or location, but instead incorporate the biochemical mechanisms in an abstract way. Furthermore, these models do not attempt to represent biomechanical mechanisms of remodeling. Instead, the authors relate remodeling cell populations to the change in bone mass, or bone volume, thus allowing a way to track the physical effects of resorption and formation. This approach was used in \cite{lio2012} to model osteomyelitis. Others, \emph{e.g.}, \cite{buckwright2004,buckwright2005}, have modeled only the resorption phase of remodeling. These models do not track remodeling cell populations, but rather cellular activity represented by enzyme reaction equations. All of these models, as well as most others, incorporate the actions of the most prominent signaling pathways involved in bone remodeling, as well as various other autocrine and paracrine signaling pathways depending on the specific considerations.

The models presented in \cite{buenzli2012b,buenzli2011,pivonka2008,pivonka2010b} are also cell population models that describe bone remodeling cell dynamics as controlled by autocrine and paracrine signaling. These models use a biochemically explicit approach, in that the model equations contain terms corresponding to interactions of remodeling cells with specific cytokines. Another distinction is that while the models represented in \cite{komarova2005,komarova2003,buckwright2004,buckwright2005} represent remodeling on the scale of an individual basic multicellular unit (BMU), the models in \cite{buenzli2012a,lemaire2004,pivkom2010a,pivonka2008} average over a volume of bone. By a BMU we mean a localized collective of bone cells working as an organized unit to carry out a single, complete cycle of the remodeling process, see \emph{e.g.}, \cite{pobb}. We note that the models found in [1, 5, 6, 18, 44, 45] consider remodeling of individual BMUs, but in a spatially explicit context, which is beyond the scope of this paper. However, the model developed herein can be easily modified to obtain a spatially explicit model.

The complexity of the skeletal system and its interactions with the rest of the body limits the ability of a single model to capture all of the relevant biological, biochemical, and biophysical mechanisms of remodeling on all scales simultaneously. Moreover, parameterization of a given model is difficult, given the dependency on accuracy and availability of data. Here we introduce a novel mathematical model for bone remodeling that maintains those fundamental mechanisms captured in previous models, while incorporating biological aspects of bone remodeling that have not previously been considered. In particular, this model includes osteocytes, their biochemical processes, and their interactions with other bone remodeling cells, and thus represents a significant advance to the field.
\section*{Materials and Methods}
\subsection*{Mathematical Modeling}

In developing a new mathematical model for bone remodeling we have chosen to employ the biochemical systems analysis formalism used in \emph{e.g.} \cite{ayati2010,graham2012,komarova2005,komarova2003,ryser2010,ryser2009}, but extended to incorporate further biological detail. We chose this approach in anticipation of the type of data to which we may eventually have access, and the types of questions which are of concern to us. For example, explicit chemical data are not obtained in current clinical practice. Our approach does represent many of the mechanisms contained in \cite{buenzli2012a,lemaire2004,pivonka2008,pivonka2010b}, albeit implicitly. In particular, we incorporate the actions of an important signaling pathway based on the specific molecules: receptor activator of nuclear factor $\kappa$-B (RANK); and osteoprotegerin (OPG) \cite{bell2003,hofbauer2001,hofbauer2004a,hofbauer2004b,martin2004,simonset1997}. These cytokines, plus the RANK ligand, form a pathway commonly known as the RANK/RANKL/OPG pathway. We also incorporate the actions of growth factors such as transforming growth factor $\beta$ (TGF-$\beta $) \cite{cohen1997,janssens2005}, and other cytokines on bone remodeling cells. It is well known that RANKL is a key cytokine in the differentiation process of osteoclast cells, while OPG, which is produced by differentiated osteoblastic cells, has been shown to function as an inhibitory factor for osteoclastogenesis \cite{pobb}.

We have developed a cell population model for osteocyte-induced targeted bone remodeling. This model consists of the osteocyte, pre-osteoblast, osteoblast, and osteoclast cell populations; the interactions of these cells with one another, and through the power law formalism the autocrine and paracrine signaling among these cells. Figure 2 shows the equations that make up the model, these are described in detail below. In this model we explicitly include the population of pre-osteoblasts to emphasize the switch of cells in the osteoblast lineage from being osteoclastogenic, \emph{i.e.}, osteoclast generating, to being osteogenic, \emph{i.e.}, bone forming, see \emph{e.g.}, \cite{boyle2003,gori2000}. Specifically, there is a clear distinction between the signaling behaviors of pre-osteoblasts and osteoblasts that is important for the considerations taken up in this work. On the other hand, for the purposes we have in mind there is no similar \emph{a priori} reason to explicitly represent a population of pre-osteoclasts.

A general assumption is that there is a large pool of mesenchymal stem cells available to differentiate into pre-osteoblasts \cite{pivonka2008}. Similarly, we assume there is a large pool of osteoclast progenitor cells which are available to differentiate to fully committed mature osteoclasts \cite{pivonka2008}. Such cell differentiation is determined by autocrine and paracrine signaling discussed in more detail below, see also \cite{kular2012}. We assume that some percentage of pre-osteoblasts differentiate under the influence of autocrine and paracrine signaling while some percentage undergo apoptosis. We also assume that some percentage of mature osteoblasts will undergo apoptosis, and some percentage of osteoblasts will become embedded in the bone matrix as osteocytes . Figure 1 outlines the assumptions that we make to construct the mathematical model. The details of the assumptions and how they influence the development of a mathematical model is described in detail below.

In the following we denote by $S(t)$, or simply $S$, the osteocyte cell population at given time $t$. Sclerostin is produced by osteocytes and inhibits the Wnt/$\beta $-catenin pathway, \cite{bonewald2011,kular2012}. Wnt is known to promote osteoblastic proliferation and differentiation \cite{neve2012}. We incorporate the effects of sclerostin and the Wnt/$\beta $-catenin pathway into the mathematical model through a term of the form ${\left(1-\frac{S}{K_S}\right)}_{+}$, where $(x)_{+} =max(x ,0)$, and $K_S$ is a parameter that describes the relation between osteocyte apoptosis and decrease in sclerostin inhibition.  The idea is that, for a threshold level $K_S$ of osteocytes, there is sufficient sclerostin production to inhibit local Wnt signaling. When osteocytes die, the sclerostin level decreases, \cite{bonewald2011,kular2012}. This releases osteoblast precursor cells from Wnt inhibition, thereby initiating a cycle of targeted bone remodeling. We note that the term $\cs$ is a dimensionless quantity that is used to modify the standard biochemical systems analysis formalism in order to accurately capture the action of sclerostin signaling as previously described. It is also important to note that in the biochemical systems analysis formalism the basic relations used to represent signaling are nonlinear. However, it may be the case that the exponents used take on numerical values equal or close to unity. This should not be misconstrued as an \emph{a priori} assumption of linearity within the model.

Equation (1) in Figure 2 describes the dynamics of the osteocyte cell population. This equation simply states that osteocytes are mature osteoblasts that become embedded in extra cellular matrix at a given rate ${\alpha }_{{\rm 1}}$. We further note that there is no death term for osteocytes in equation (1). This is due to our assumption that, over the time scale of a single event of targeted remodeling considered here, the most significant influence on osteocyte apoptosis is the initial biomechanical action that begins remodeling \cite{cardoso2009,heino2009}. This appears in the mathematical model as an initial condition.

The pre-osteoblast cell population at a time $t$ is denoted by $P{\rm (}t{\rm )}$, or simply $P$. Pre-osteoblasts are differentiated mesenchymal stem cells. We assume that this differentiation is controlled by osteocytes through the sclerostin, Wnt/$\beta $-catenin pathways, and various growth factors. The effectiveness of sclerostin regulations of the differentiation of mesenchymal stem cells to become pre-osteolbasts is represented mathematically by ${\left({\rm 1-}\frac{S}{K_S}\right)}^{g_{{\rm 22}}}_{{\rm +}}$. Where $g_{{\rm 22}}$ is a dimensionless parameter. Thus when osteocytes undergo apoptosis due to microdamage, local mesenchymal stem cells differentiate to pre-osteoblasts due to the resulting signaling. Moreover, the pre-osteoblasts are free to proliferate and differentiate to mature osteoblasts since they have been released from Wnt inhibition. Equation (2) in Figure 2 describes the dynamics of the pre-osteoblast cell population. This equation states that pre-osteoblasts are differentiated from a large pool of stem cells at a rate ${\alpha }_{{\rm 2}}$ in response to signaling molecules produced by osteocytes, that pre-osteoblasts proliferate at a rate ${\alpha }_{{\rm 3}}$ under the influence of autocrine signaling provided this is not inhibited by sclerostin. Furthermore, pre-osteoblasts differentiate to become mature osteoblasts at a rate ${\beta }_{{\rm 1}}$. This is under the influence of autocrine and osteoclast regulated paracrine signaling. Finally, pre-osteoblasts undergo apoptosis at a rate $\delta $.

The osteoblast cell population at a time $t$ is denoted by $B{\rm (}t{\rm )}$, or simply $B$. Equation (3) in Figure 2 describes the dynamics of the mature osteoblast cell population. This equation states that osteoblasts are differentiated pre-osteoblasts, that osteoblasts undergo apoptosis, and that some osteoblasts are embedded in the extra cellular matrix during formation to become osteocytes. Notice that the term ${\alpha }_{{\rm 1}}B^{g_{{\rm 31}}}{\left({\rm 1-}\frac{S}{K_S}\right)}_{{\rm +}}$ also appears in equation (1) of Figure 2, representing the embedding of osteoblasts that become osteocytes, and the term ${\beta }_{{\rm 1}}P^{f_{{\rm 12}}}C^{f_{{\rm 14}}}$, also appears in equation (2) of Figure 2, which corresponds to the differentiation of pre-osteoblasts to become mature osteoblasts. Here the parameter $f_{{\rm 12}}$ describes the pre-osteoblast autocrine signaling. The parameter $f_{{\rm 14}}$ represents the effects of osteoclast derived paracrine signaling on pre-osteoblasts. This could represent, for example, the effects of TGF-$\beta $ on pre-osteoblasts as described for example in \cite{pivonka2008}.

The osteoclast cell population at a time $t$ is denoted by $C{\rm (}t{\rm )}$, or simply $C$. The equation (4) in Figure 2 describes the dynamics of the osteoclast cell population. This equation states that mature osteoclasts come from the differentiation of a large pool pre-osteoclasts at a rate ${\alpha }_{{\rm 4}}$. This differentiation is influenced essentially by the RANK/RANKL/OPG pathway. Thus the term $S^{g_{{\rm 41}}}P^{g_{{\rm 42}}}{\left(\varepsilon {\rm +}B\right)}^{g_{{\rm 43}}}{\left({\rm 1-}\frac{S}{K_S}\right)}^{g_{{\rm 44}}}_{{\rm +}}$, describes the effects of this pathway. The dimensionless parameter $g_{{\rm 44}}$ models the effectiveness of sclerostin regulation of osteoclastogensis. One novel feature of this model is that we have included osteocytes as a source of RANKL as discussed in recent literature \cite{bonewald2011,heino2009,kular2012,neve2012,xiong2011}. The parameter $g_{{\rm 41}}$ represents the effect of osteocyte derived RANKL signaling on osteoclastogenesis. We also retain pre-osteoblast derived RANKL signaling via the parameter $g_{{\rm 42}}$. While we have explicitly included the pre-osteoblast cell population dynamics, we have neglected to explicitly include a dynamic cell population for pre-osteoclasts. This is based on a simplifying assumption that, over the relevant time span, there is a steady pool of circulating  pre-osteocalsts that may be recruited for differentiation to active osteoclasts. This simplifying assumption is common the literature on mathematical modeling of bone remodeling, \emph{e.g.}, \cite{ayati2010,buenzli2012b,buenzli2012a,buenzli2011,komarova2005,komarova2003,lemaire2004}.

We note that either of these, $g_{{\rm 41}}$ or $g_{{\rm 42}}$, could take on the value $0$. Thus advances in the understanding of the relative role of osteoblasts and osteocytes in RANKL production would result in parameter changes in our model, but not a change in the model structure itself.

The term ${\left(\varepsilon {\rm +}B\right)}^{g_{{\rm 43}}}$ represents the effect of OPG acting as a decoy receptor for RANKL. Typically the parameter $g_{{\rm 43}}$ takes on negative values, and since $B$ has $0$ as a steady state value, we add a sufficiently small number $\varepsilon $ to avoid dividing by zero. This represents the factor of production when $B{\rm =0}$. We also have a term for the osteoclast cell death at a rate ${\beta }_{{\rm 3}}$.

We denote by $z{\rm (}t{\rm )}$, or simply $z$, the bone mass at a given time $t$. We follow \cite{komarova2003,pivonka2008} to develop an equation for the change of bone mass (or bone volume depending on the scaling) over time. The rate of change of bone mass has the form
\[ \frac{dz}{dt} = \mbox{formation - resorption}. \]
 As in \emph{e.g.}, \cite{pivonka2008}, we assume that the amount of bone being resorbed is proportional to the osteoclast population, while the amount of bone being formed is proportional to the osteoblast population.

We can summarize the mathematical model with the following verbal description:

\begin{enumerate}
\item  \textbf{change in osteocytes} =  increase due to  embedded osteoblasts

\item  \textbf{change in pre-osteoblasts} = increase due to differentiation of stromal cells (released from sclerostin or exposure to growth factors) + proliferation of pre-osteoblasts (autocrine signaling of Wnt and growth factors) -- differentiation to osteoblasts (growth factors) - apoptosis

\item  \textbf{change in osteoblasts} =  increase due to differentiation of pre-osteoblasts (growth factors) -- apoptosis -- embedding as osteocytes

\item  \textbf{change in osteoclasts} = increase due to differentiation of pre-osteoclasts (due to RANKL and limited by OPG) - apoptosis

\item  \textbf{change in bone mass} = increase due to activity of osteoblasts -- activity of osteoclasts
\end{enumerate}

The corresponding full mathematical model is the set of equations shown in Figure 2. The definitions of the variables and parameters appearing in the model equations of Figure 2 is summarized in table 1.

\subsection*{Initial Conditions}

The initial conditions for the mathematical model are given by $S(0)=K_{S}-\rho$, where $\rho >0$ is a constant that represents osteocyte apoptosis, which corresponds to a drop in sclerostin levels, $P(0)=B(0)=C(0)=0$, and $z(0)=100$. Observe that the system of model equations (1)-(4) has as a steady state, the values $S_{ss}=K_{S}$, $P_{ss}=0$, $B_{ss}=0$, and $C_{ss}=0$. The significance being that this steady state allows one to model the initiation of an event of targeted bone remodeling. In particular, our model does not assume the presence of a population of committed pre-osteoblasts or active osteoclasts in order to begin a remodeling cycle. This is in contrast with previously published mathematical models of bone remodeling which require either a population of active osteoclasts, or a population of pre-osteoblasts to initiate dynamics corresponding to physiological bone remodeling. However, it should be noted that physiological bone remodeling also involves the movement of bone lining cells away from the bone surface. This aspect of bone remodeling is complicated and not yet fully understood. The steady state value for bone volume, denoted $z_{ss}$ depends significantly on the parameter values. For instance, if the parameter values are such that there is an imbalance of osteoblasts and osteoclasts, in favor of osteoblasts, then there will be over remodeling.

\subsection*{Numerical Simulations}

The mathematical model is a system of coupled nonlinear ordinary differential equations (ODEs). To use the mathematical model to simulate bone remodeling we must solve this system along with the specified initial conditions. In order to do so we use the MATLAB software for scientific computing. In particular, we use the MATLAB ODE suite \cite{shampine1997}, and the program ode23s to integrate the problem (1)-(5) with the above initial conditions. For a further discussion of the numerical techniques underlying the codes in the MATLAB ODE suite see \cite{shampine2003,shampine1994}.

\subsection*{Parameters}

Table 2 lists the values of the parameters in the model equations (1)-(5) that we have taken as baseline values. These particular values are chosen so that a cycle of normal, targeted bone remodeling is completed, \emph{i.e.}, bone cell populations reach a steady state value, after approximately one hundred days.

% Results and Discussion can be combined.
\section*{Results}

\subsection*{Baseline Case}

Figure 3 shows the simulation results for bone cell populations during a single cycle of normal targeted bone remodeling. Here, normal targeted bone remodeling is defined to be a complete remodeling cycle that takes place over a period of one hundred days, in which the amount of new bone formed is equal to the amount of old or damaged bone resorbed. One characteristic of normal bone remodeling that manifests itself quantitatively is that the steady state value of bone volume, here and below denoted by $z_{ss}$, is equal to one hundred percent. The simulations shown in Figure 3 result from solving the model equations (1)-(5) with the above initial conditions, and parameter values shown in table 2. Figure 3(a) shows the dynamics of the osteocyte cell population during an event of targeted bone remodeling. Initially there is a decrease from the steady state value $K_S$ in the osteocyte cell population. This corresponds to local osteocyte apoptosis, and results in a decrease in local sclerostin levels. This releases stromal cells from sclerostin inhibition, allows for Wnt signalling, and results in proliferation and differentiation of pre-osteoblast cells, Figure 3(b). There follows an increase in osteoblast cell number due to differentiation of pre-osteoblasts, and differentiation of osteoclast pre-cursors to mature osteoclasts, Figure 3(c) and Figure 3(d). Remodeling ceases once the local osteocyte cell population is replenished, and the osteocyte cell network is reestablished, returning sclerostin expression back to sufficient levels thereby inhibiting bone turnover.

Figure 4 shows the dynamics of bone volume during a single event of targeted bone remodeling. An increase in the osteoclast cell numbers results in bone resorption and the decrease in bone volume, while increases in osteoblast cell number results in bone formation. In this case bone turnover is completely balanced in the sense that the amount of new bone formed equals the amount resorbed. Mathematically this corresponds to a steady state value of bone mass, $z_{ss}{\rm =100\%}$.

\subsection*{Role of Osteocyte RANKL Production}

At the scale of a BMU, bone cells are easily affected by changes in the microenvironment. Thus we expect that small changes in the model parameter values could significantly impact bone turnover during remodeling. In this section we study the influence of expression of RANKL by osteocytes on targeted bone remodeling. In particular, we are interested in the question of whether local RANKL expression by pre-osteoblasts, or by osteocytes has the greater impact. Following \cite{pivonka2008}, we take as a control parameter the change in bone volume. This provides a measurable quantity by which we can assess the reasonableness of the mathematical model. If $z_{ss}{\rm >100}$ then there is over-remodeling. If $z_{ss}{\rm <100}$ then there is under-remodeling.

That RANKL/OPG expression by cells in the osteoblastic lineage is fundamental for the activation and control of bone remodeling is by now well established. Until recently, the paradigm was that pre-osteoblasts serve as the primary source for RANKL and active osteoblasts serve as the primary source of OPG \cite{gori2000}. More recent research highlights that osteocytes also express RANKL at significant levels, and have a strong influence on the activation and regulation of remodeling \cite{bonewald2011,guiliani2012,heino2009,kular2012,neve2012,xiong2011}. To explore this we vary the parameters $g_{{\rm 41}}$ and $g_{{\rm 42}}$ in equation (4) over the range from zero to two. These parameters correspond to the effectiveness of the expression of RANKL by osteocytes, and pre-osteoblasts respectively. By varying these parameter values simultaneously we gain insight into how RANKL expression by osteocytes and pre-osteoblasts impacts the regulation of a cycle of targeted bone remodeling. We note that the system is very sensitive to changes in the effectiveness of RANKL expression. In particular, small perturbations to the amount of RANKL can lead to pathological remodeling. First we compute $z_{ss}$ as a simultaneous function of $g_{{\rm 41}}$ and $g_{{\rm 42}}$, and observe that values of $g_{{\rm 41}}$ and $g_{{\rm 42}}$ that are simultaneously large, in this case ${\rm >1}$, result in physiologically unrealistic over resorption (under remodeling). Next we compute $z_{ss}$ as a function of each of $g_{{\rm 41}}$ and $g_{{\rm 42}}$ individually with the other fixed. We observe that it is possible to have values such that either $g_{{\rm 41}}{\rm >}g_{{\rm 42}}$, or $g_{{\rm 41}}{\rm <}g_{{\rm 42}}$ and resulting in $z_{ss}\approx {\rm 100\%}$. Table 3 lists some specific values of $g_{{\rm 41}}$, and $g_{{\rm 42}}$ that result in normal targeted bone remodeling. Thus, given cell population data, this model can be used, with parameter fitting, to determine whether osteocytes or pre-osteoblasts provide the dominant RANKL source during targeted bone remodeling. This allows for more thorough testing of hypotheses such as in \cite{xiong2011}, which suggests that RANKL produced by osteoblast precursors contribute little to adult bone remodeling.

\subsection*{Signaling of Osteocytes and Pre-osteoblasts}

In this section we explore the behavior of the system as it is affected by changes in the effectiveness of osteocyte paracrine, and pre-osteoblast autocrine signaling. In particular, we regard  how these actions influence the differentiation of stromal cells into pre-osteoblasts, and the proliferation of pre-osteoblast cells. These aspects of BMU remodeling are strongly influenced by sclerostin signaling. While previous mathematical models have included pre-osteoblast proliferation, explicit inclusion of osteocyte signaling, and the effects sclerostin, has been outside the scope of those works and thus is another feature of novelty in this work.

We compute the steady state bone volume, $z_{ss}$, as a simultaneous function of the effectiveness of osteocyte paracrine signaling on stromal cell differentiation, $g_{{\rm 21}}$, and pre-osteoblast autocrine signaling for pre-osteoblast proliferation, $g_{{\rm 32}}$. The results are shown in Figure 5, Figure 6, and Figure 7. The reason to consider these parameters simultaneously is that these terms both influence the ``production'' of pre-osteoblasts in (2). First, we observe that the parameter $g_{{\rm 21}}$ is considerably more sensitive to variation than $g_{{\rm 32}}$, this is seen in Figure 5 and in a comparison between Figure 6 and Figure 7. In the context of the model, this suggests that the influence of osteocyte signaling, and sclerostin specifically, on stromal cell differentiation into pre-osteoblasts is key to the model. In fact, we can obtain a steady state value for bone volume, $z_{ss}$, corresponding to normal bone remodeling for a relatively wide range of $g_{{\rm 32}}$ values, provided $g_{{\rm 21}}$ is at or near baseline value. The significance of the role of osteocyte signaling is particularly strong with regard to the initiation of a cycle of targeted remodeling, since as pointed out above, the steady state values for all cell populations, except osteocytes, is zero. That is, we do not have a presence of cells fully committed to either the osteoblast or osteoclast lineage when there is no active remodeling. This is a point of departure with previous mathematical models where positive steady state values of mature osteoblasts and osteoclasts occur.

\subsection*{Application to Study Anti-sclerostin Drugs}

We can apply the model system to study the effects of using an anti-sclerostin drug. Many drugs for treating bone disease, such as biphosphonates, are designed to inhibit resorption, and to decrease the loss of bone mass. On the other hand, anti-sclerostin drugs purport to promote bone formation, thereby increasing bone mass \cite{lewiecki2011,lim2012}. Sclerostin inhibits Wnt/$\beta $-catenin, so anti-sclerostin would promote osteoblast differentiation. In addition, Wnt signaling increases OPG expression, which inhibits osteoclasts, so anti-sclerostin would be expected to increase OPG release too. Sclerostin has been shown to downregulate OPG and increase levels of RANKL. By modifying the parameters $g_{{\rm 22}}$ and $g_{{\rm 44}}$ which represent the effectiveness of sclerostin regulation of osteoblastogenesis and osteoclastogenesis respectively, albeit in an abstract manner, we can study the effects of an anti-sclerostin agent on bone remodeling in a situation where there is known to be an imbalance in resorption/formation (\emph{e.g.}, abnormal osteoclast activity).

To model the inclusion of an anti-sclersotin drug we proceed as follows. In general, the parameters $g_{ 22}$ and $g_{ 44}$ now become time dependent, that is $g_{22}(t )$ and $g_{44}(t)$. The form of the time dependence is determined by the frequency of dosing, the half life of the drug, and potentially the method of drug delivery. We note that since $g_{22}$ and $g_{44}$ are exponents in a power law approximation they must be dimensionless quantities, this remains the case even when they are made to depend on time. We take the function $g_{22}(t )$ to be a constant value $G_{22}$ over a specific time interval $t_{{\rm 1}}$ to $t_{{\rm 2}}$, and the baseline value otherwise. We take the function $g_{{\rm 44}}{\rm (}t{\rm )}$ to be a constant $G_{{\rm 44}}$ over another time interval $t_{{\rm 3}}$ to $t_{{\rm 4}}$, and the baseline value otherwise. The constants $G_{22}$ an $G_{44}$ represent the perturbation of the baseline parameter values for $g_{{\rm 22}}$ and $g_{{\rm 44}}$ respectively. In order to get an increase in bone formation we must have $G_{{\rm 2}{\rm 2}}$ be less than the baseline value, and $G_{{\rm 44}}$ be greater than the baseline value. The time intervals $t_{{\rm 1}}$ to $t_{{\rm 2}}$, and $t_{{\rm 3}}$ to $t_{4}$ represent the period of activity of the given dose of the drug. The time interval $t_{1}$ to $t_{2}$ is the time over which the drug effects Wnt signaling. The time interval $t_{3}$ to $t_{4}$ is the time over which the drug effects OPG signaling. We note that these two time intervals may be the same.

Figure 8 shows the change in bone mass as a function of time in the case of simulated pathological remodeling. Here we simulate the results of a hypothetical bone disease that results in over resorption. This is modeled by increasing the value of $\alpha_{4}$ from its baseline value as listed in table 2 to a value of ${\alpha }_{{\rm 4}}{\rm =0.11}$. Figure 8 shows the bone mass as a function of time under conditions where there is over resorption and no treatment, that is all parameters other than $\alpha_{4}$ are set to baseline values, and, in particular we do not yet modify $g_{{\rm 22}}$ and $g_{{\rm 44}}$.

Next, we simulate the treatment of the pathological remodeling, such as is show in Figure 8, via an anti-sclerostin agent. Figure  9 shows the results obtained by including treatment with an anti-sclerostin drug, implemented by modifying the parameters $g_{{\rm 22}}$ and $g_{{\rm 44}}$ as described above. Specifically, all the parameters are the same as those used to obtain the results show in 8, but now we modify $g_{{\rm 22}}$ and $g_{{\rm 44}}$ to be the functions $g_{{\rm 22}}{\rm (}t{\rm )}$ and $g_{{\rm 44}}{\rm (}t{\rm )}$, previously defined, to model the effects of treatment with an anti-sclerostin antibody. We observe a dose dependent increase in bone mass as described in \cite{lewiecki2011,lim2012}.

There are several notable features regarding the treatment of bone loss using an anti-sclerostin antibody within the context of this model. One is that the time of the dosing matters. The closer to initial stages of osteoclastogenesis that the dose is given, the better the response of increased bone formation. This is particularly the case with respect to OPG signaling related parameter $g_{{\rm 44}}$. Another feature we observed is that the parameter $g_{{\rm 22}}$, which corresponds primarily to Wnt signaling, is sensitive to the perturbations that model dosing with an anti-sclerostin drug. That is, it takes only a small decrease in the value over a short period of time for a marked increase in bone formation.

\section*{Discussion}

We have developed a cell population mathematical model in the form of a system of nonlinear ordinary differential equations for the principal bone cells involved in local targeted bone remodeling. The primary goal for this work was to include features of major focus in some of the most recent experimental research on bone remodeling at the cellular level. In particular we have aimed to present a model that extends and compliments current mathematical models and allows for a theoretical testing of hypotheses put forth in recent experimental work that is difficult to test \textit{in vivo} or in clinical settings.

The most prominent consideration in the development of this model, that differs significantly from previous work, has been the explicit inclusion of osteocyte cell populations, and the effects of osteocyte signaling on bone turnover during targeted bone remodeling. First, in contrast to previous mathematical models that required perturbation away from a positive steady state to begin a remodeling cycle, our model allows for a more natural (physiological) initiation of a cycle of targeted bone remodeling. Second, the steady state values for forming or resorbing bone cells is in better agreement with what is physiologically expected, since the presence of mature osteoclasts and osteoblasts is not expected in the absence of a remodeling BMU. Finally, the approach taken here allows for a natural representation of the action of sclerostin which inhibits Wnt signaling and pre-osteoblast proliferation, and may also promote osteoclast activity.

We have used, following \cite{pivonka2008}, the steady state bone volume as a control to examine aspects of targeted bone remodeling such as osteocyte RANKL expression, initiation of targeted bone remodeling with osteocyte apoptosis, and the role of sclerostin that are of current interest to the larger bone remodeling community. We have shown that this model is appropriate for determining whether osteocyte or pre-osteoblast RANKL expression plays the more prominent role in osteoclastogenesis for adult bone remodeling. Furthermore, we have applied this model to study the effects of an anti-sclerostin drug, such as an anti-sclerostin monoclonal antibody described in \cite{lewiecki2011}, on bone turnover.

One important feature of bone remodeling that is beyond the scope of the model presented in this work is spatial bone remodeling. We point out that the main thrust of spatially explicit mathematical models of bone remodeling have been spatial extensions of purely temporal models, see \emph{e.g.}, \cite{ayati2010,ryser2010}. That is, the ordinary differential equations used to model local bone remodeling are extended to partial differential equations.

However, a different framework for spatial extensions based on level set methods (LSM) is also possible, see \emph{e.g.}, \cite{ayati2012,graham2012}. The models used in \cite{ayati2012,graham2012} did not include the role of osteocytes, or differentiate between precursor cells and active osteoblasts and osteoclasts. Nonetheless the works \cite{ayati2012,graham2012} provide a proof of concept for a new spatial representation of bone remodeling, and demonstrate the sort of pitfalls that might occur when spatially extending the model to scales of interest for bone marrow biopsy cores. Quite importantly, we did not see a change in the local behavior based on embedding into the LSM spatial representation; in fact local behavior translated into spatial behavior well.

Another feature, somewhat related to spatial behavior and not considered in this work are the mechanical mechanisms involved in the initiation and regulation of bone remodeling. This is primarily due to the fact that this work considers local remodeling at the scale of an individual BMU but without treating the problem of BMU steering.

Decades of experimental studies on bone remodeling reveal a dynamic process that is sensitive to multiple local and humoral factors. \textit{In silico} modeling based on this accumulated knowledge offers a means to assess systematically the effects of perturbing individual parts of the system at a rate that cannot be matched by experiments conducted \textit{in vivo}. As such, model-based predictions are particularly useful for the development of new treatments for bone disorders. This is  somewhat illustrated by results from our model, which underscore the decisive influence of osteocytes and the potential for sclerostin-targeted treatments to reduce bone loss in pathologic conditions ranging from osteoporosis to osteolytic cancers.

It has been noted that dysfunctional or apoptotic osteocytes contribute to post-menopausal, steroid-, and immobilization-induced osteoporosis (reviewed by \cite{mangolas2012}). High circulating levels of sclerostin have been associated with osteoporosis, Paget's disease, and metastatic bone disease \cite{ardawi2012,yavropoulou2012}. Intriguingly, sclerostin has also been found to be expressed by malignant plasma cells in multiple myeloma \cite{brunetti2011}. Agents targeting sclerostin are under development, representing a novel therapeutic strategy compared with such currently available options as the bisphosphonates which directly disrupt osteoclast function. Disorders of bone remodeling affect millions of individuals and are responsible for considerable morbidity and mortality, resulting billions of dollars in healthcare costs each year. Given the complexity of the cellular and biochemical processes which mediate bone remodeling, the availability of a mathematical model which can capture all of these elements is an exceedingly useful tool for understanding bone remodeling in the setting of disease. Our model, which, to our knowledge, is the first to incorporate osteocytes and sclerostin, can be used to answer questions regarding the relative impact of cell type-specific activity. Not only can the effects of existing therapeutic agents be modeled, but this model could be used to explore the effects of modulating the activity of new targets thus providing rationale for new drug development. Furthermore, this model could be extended to multiple myeloma by the incorporation of plasma cells and the resulting interactions between the bone and stromal cells.

% You may title this section "Methods" or "Models".
% "Models" is not a valid title for PLoS ONE authors. However, PLoS ONE
% authors may use "Analysis"

% Do NOT remove this, even if you are not including acknowledgments
\section*{Acknowledgments}
BPA was partially supported by the NSF under award DMS-0914514. JAM and BPA were partially supported by NIAMS grant \#1 P50 AR055533. SAH was partially supported by the PhRMA Foundation.

%\section*{References}
% The bibtex filename
\bibliography{cyteModelBone}

\section*{Figure Legends}
%\begin{figure}[!ht]
%\begin{center}
%%\includegraphics[width=4in]{figure_name.2.eps}
%\end{center}
%\caption{
%{\bf Bold the first sentence.}  Rest of figure 2  caption.  Caption
%should be left justified, as specified by the options to the caption
%package.
%}
%\label{Figure_label}
%\end{figure}

\begin{figure}[!ht]
  \begin{center}
%  % Requires \usepackage{graphicx}
  \includegraphics[width=5in]{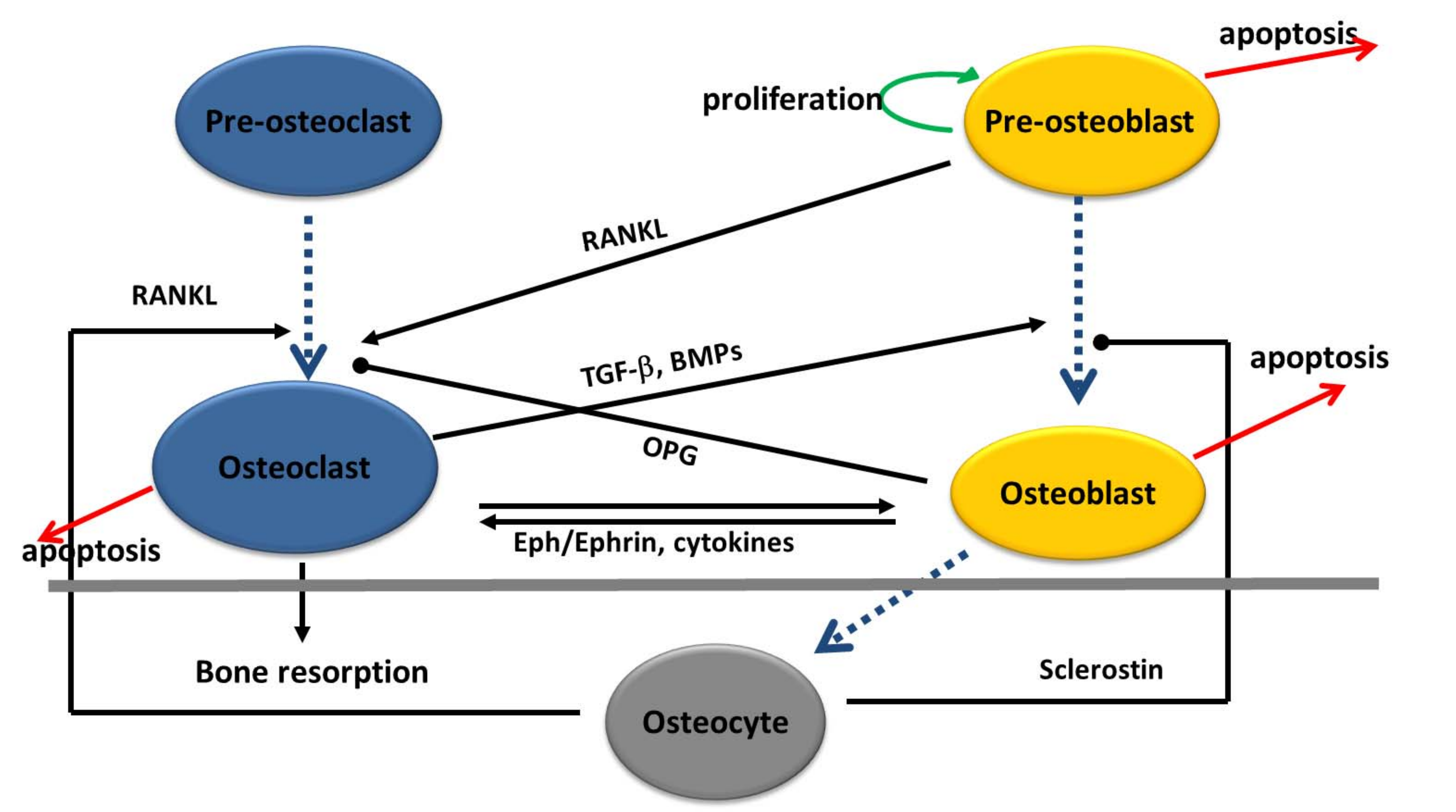}\\
  {{\bf Figure 1.} Interactions between bone cell populations. }\label{figOne}
  \end{center}
\end{figure}

\begin{figure}[!ht]
  \begin{center}
  \includegraphics[width=5in]{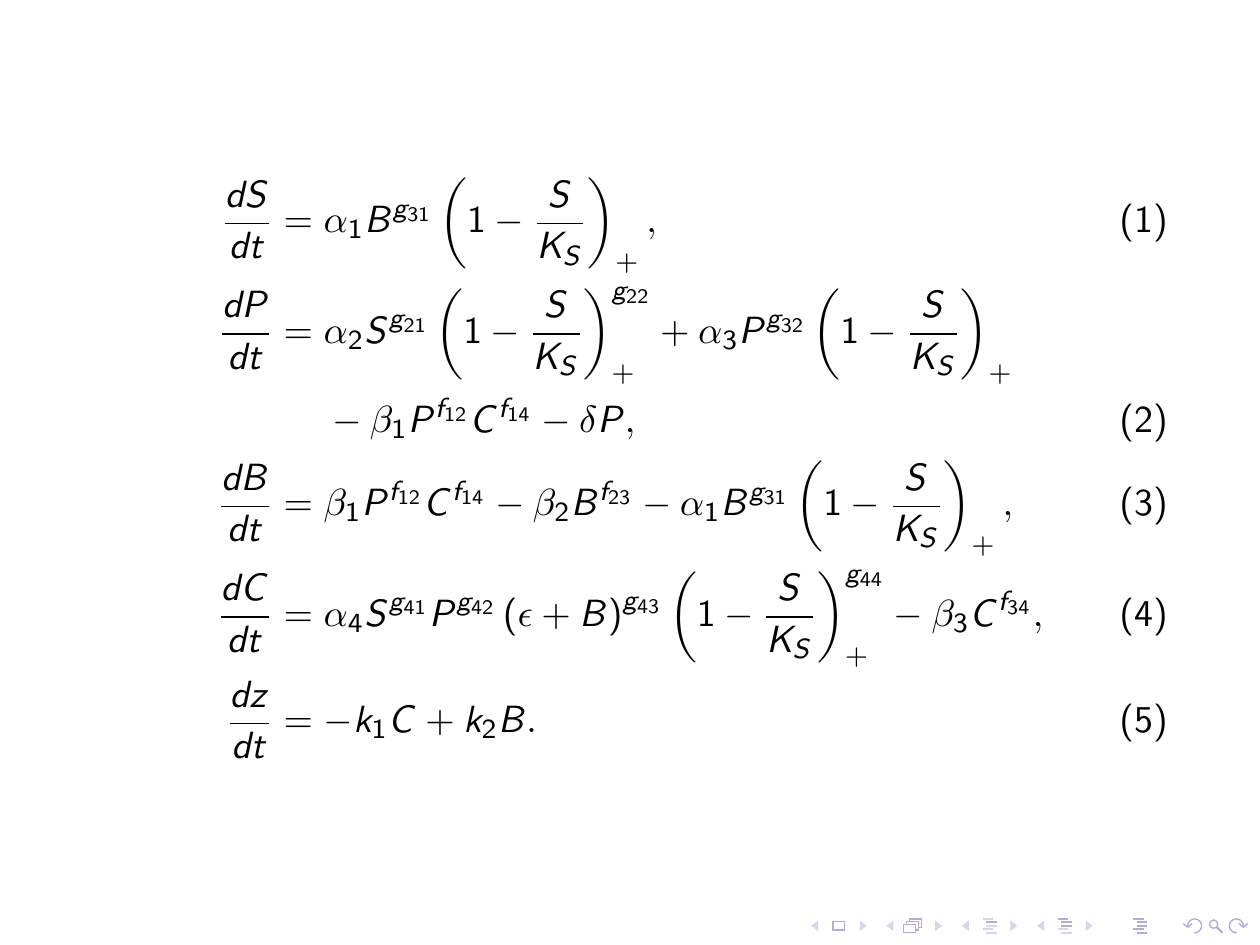}\\
  %\begin{align}
  %   \frac{dS}{dt} &= \ao B^{g_{31}}\cs, \label{eq:cytes}\\
  %   \frac{dP}{dt} &= \at S^{g_{21}}\cs^{g_{22}} + \att P^{g_{32}}\cs \nonumber \\
  %   &\ \ \ \   - \bo P^{f_{12}}C^{f_{14}} - \delta P,\label{eq:preblasts} \\
  %   \frac{dB}{dt} &= \bo P^{f_{12}}C^{f_{14}} - \bt B^{f_{23}} - \ao B^{g_{31}}\cs, \label{eq:blasts}\\
  %   \frac{dC}{dt} &=  \af S^{g_{41}} P^{g_{42}}\left(\epsilon+B\right)^{g_{43}}\cs^{g_{44}} - \btt C^{f_{34}},\label{eq:clasts} \\
  %   \frac{dz}{dt} & = -k_{1}C+k_{2}B. \label{eq:bone}
  %\end{align}
  {{\bf Figure 2.} System of ordinary differential equations constructed, using the biochemical systems analysis formalism \cite{savageau1,savageau2,savageau3,savageau1976,voit2000},  to model osteocyte-induced targeted bone remodeling. }%\label{figTwo}
  \end{center}
\end{figure}

\begin{figure}[!ht]
  \begin{center}
  \includegraphics[width=5in]{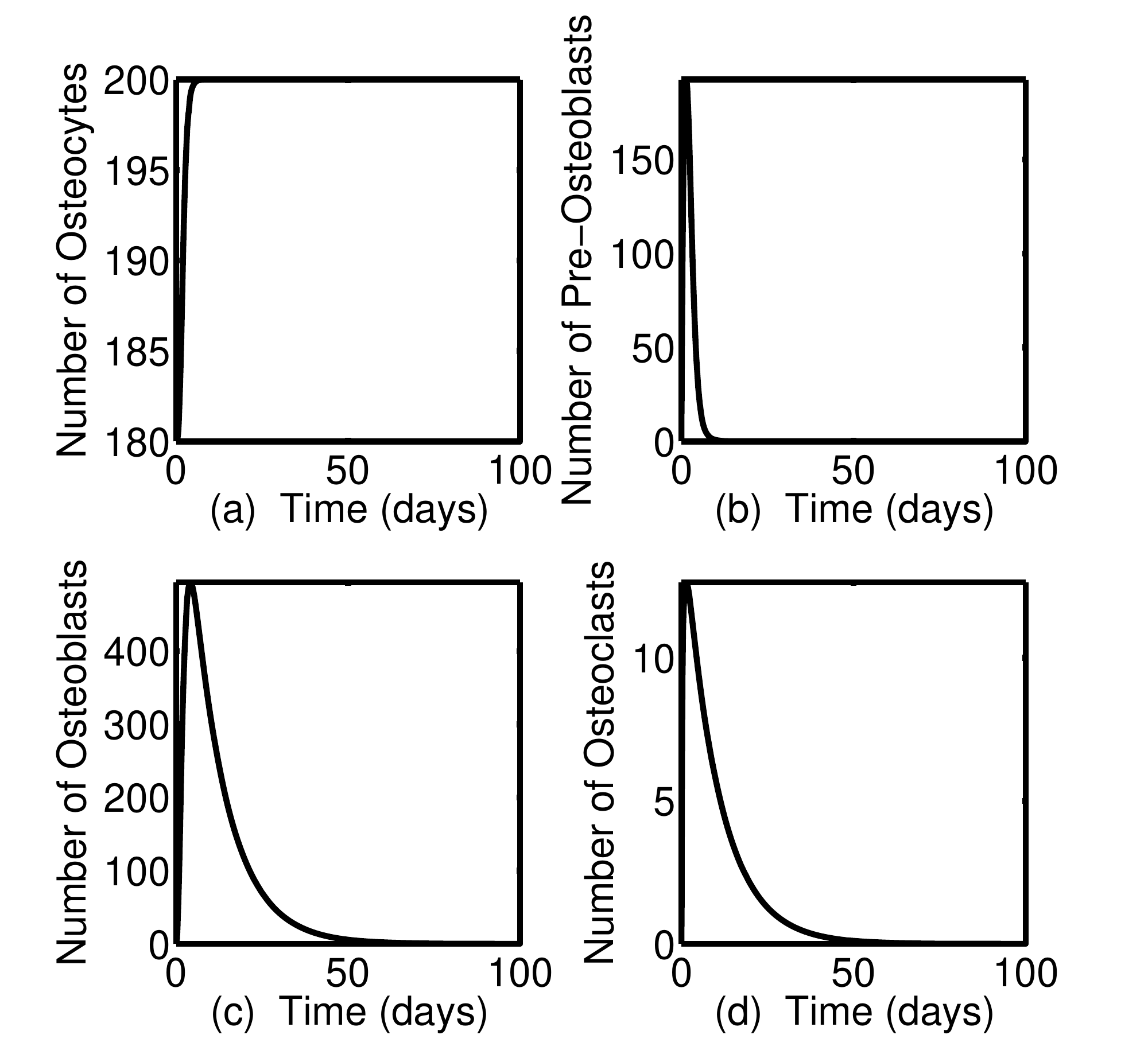}\\
  {{ \bf Figure 3.} Dynamics of bone cells during a single event of targeted bone remodeling. The dynamics of osteocyte (a), pre-osteoblast (b), osteoblast (c), and osteoclast (d) populations during an event of targeted bone remodeling. }%\label{figThree}
  \end{center}
\end{figure}

\begin{figure}[!ht]
  \begin{center}
  \includegraphics[width=4in]{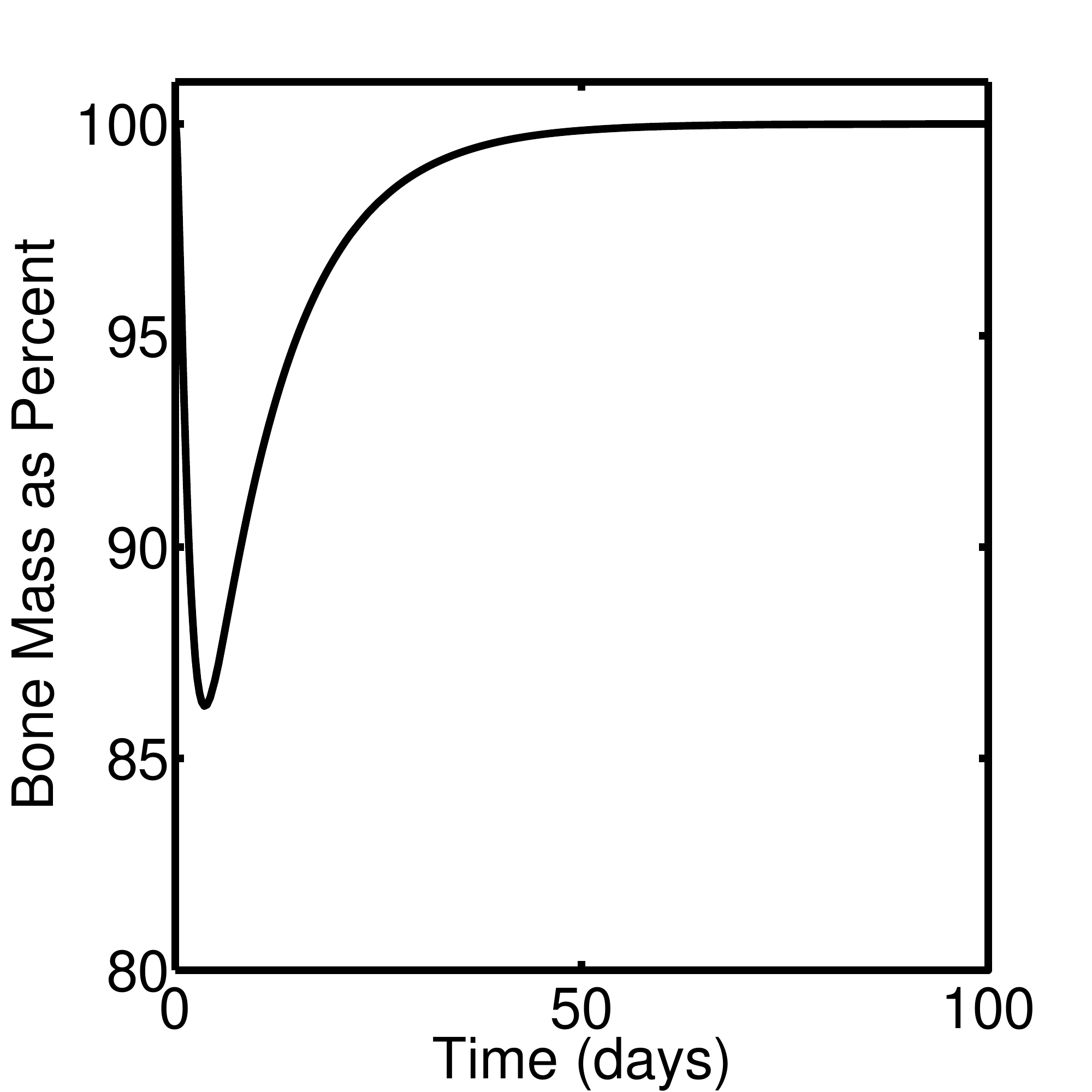}\\
  {{ \bf Figure 4.} Dynamics of bone volume during a single event of targeted bone remodeling.}%\label{figFour}
  \end{center}
\end{figure}

\begin{figure}[!ht]
  \begin{center}
  \includegraphics[width=5in]{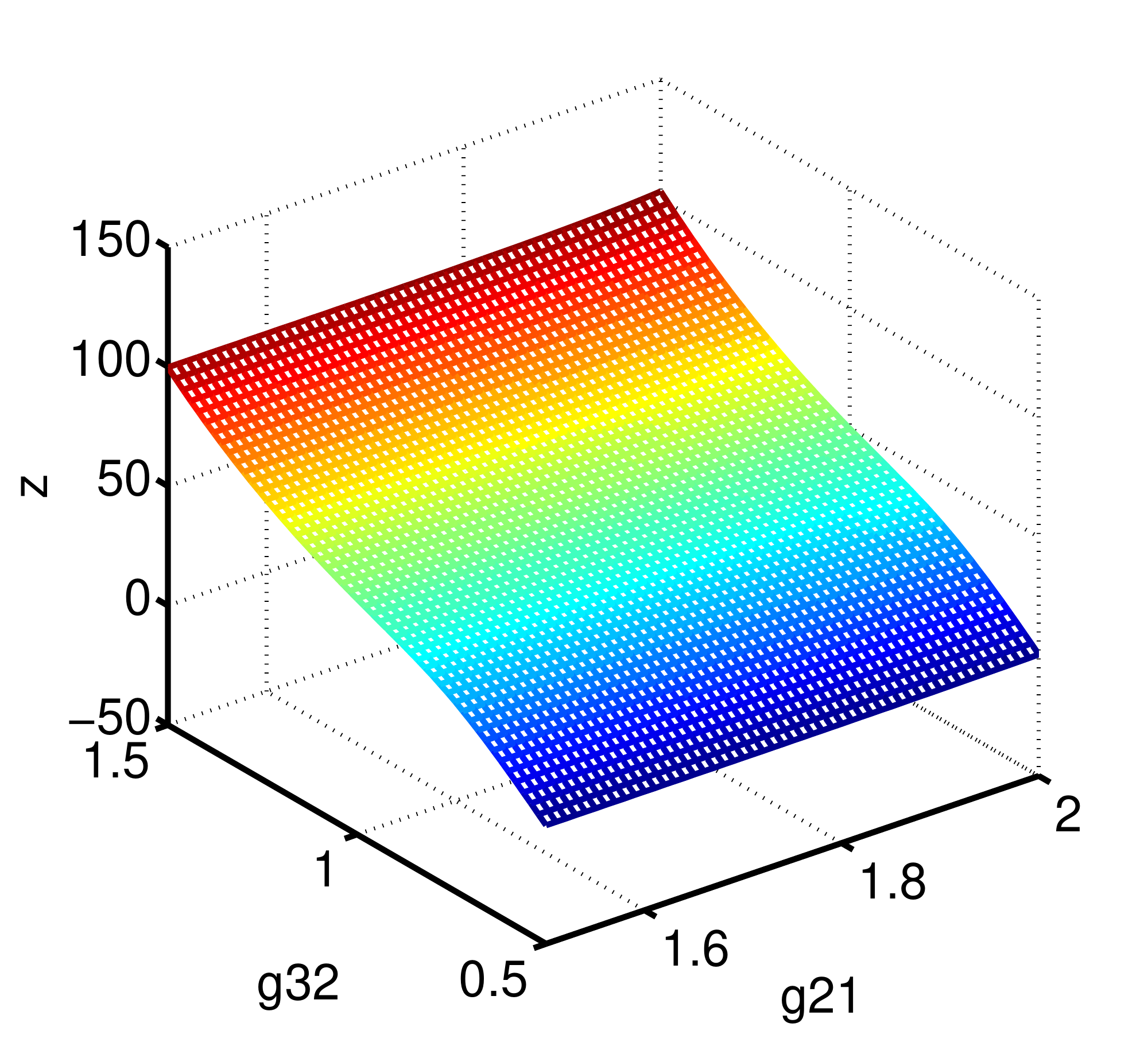}\\
  {{ \bf Figure 5.} The steady state bone volume, $z_{ss}$, as a simultaneous function of the effectiveness of osteocyte paracrine signaling on stromal cell differentiation, $g_{21}$, and pre-osteoblast autocrine signaling for pre-osteoblast proliferation, $g_{32}$. }%\label{figFive}
  \end{center}
\end{figure}

\begin{figure}[!ht]
  \begin{center}
  \includegraphics[width=5in]{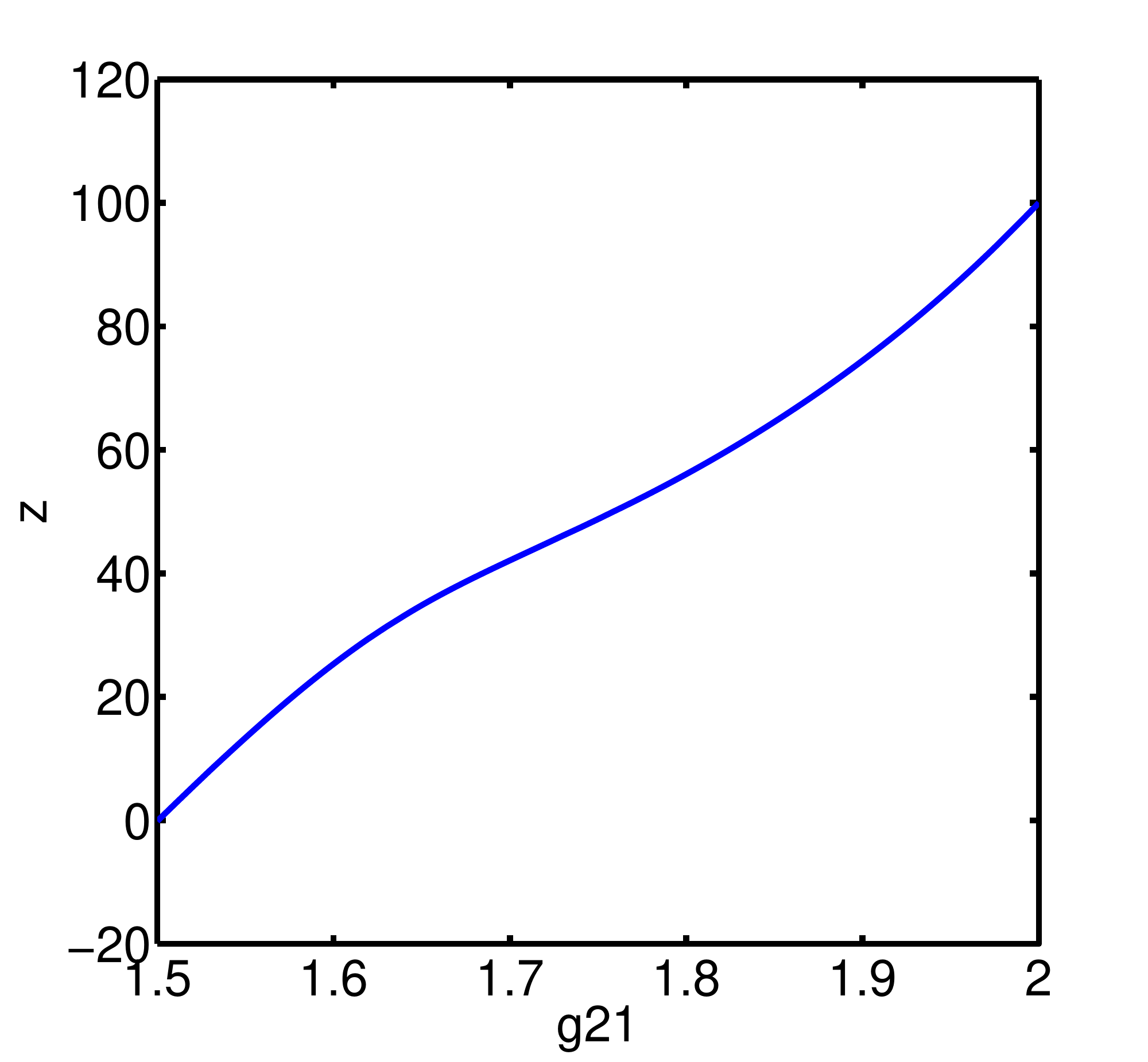}\\
  {{ \bf Figure 6.} The steady state bone volume, $z_{ss}$, computed as a function of the effectiveness of osteocyte paracrine signaling on stromal cell differentiation, $g_{21}$, with all other parameters held at baseline values.}%\label{figFive}
\end{center}
\end{figure}

\begin{figure}[!ht]
  \begin{center}
  \includegraphics[width=5in]{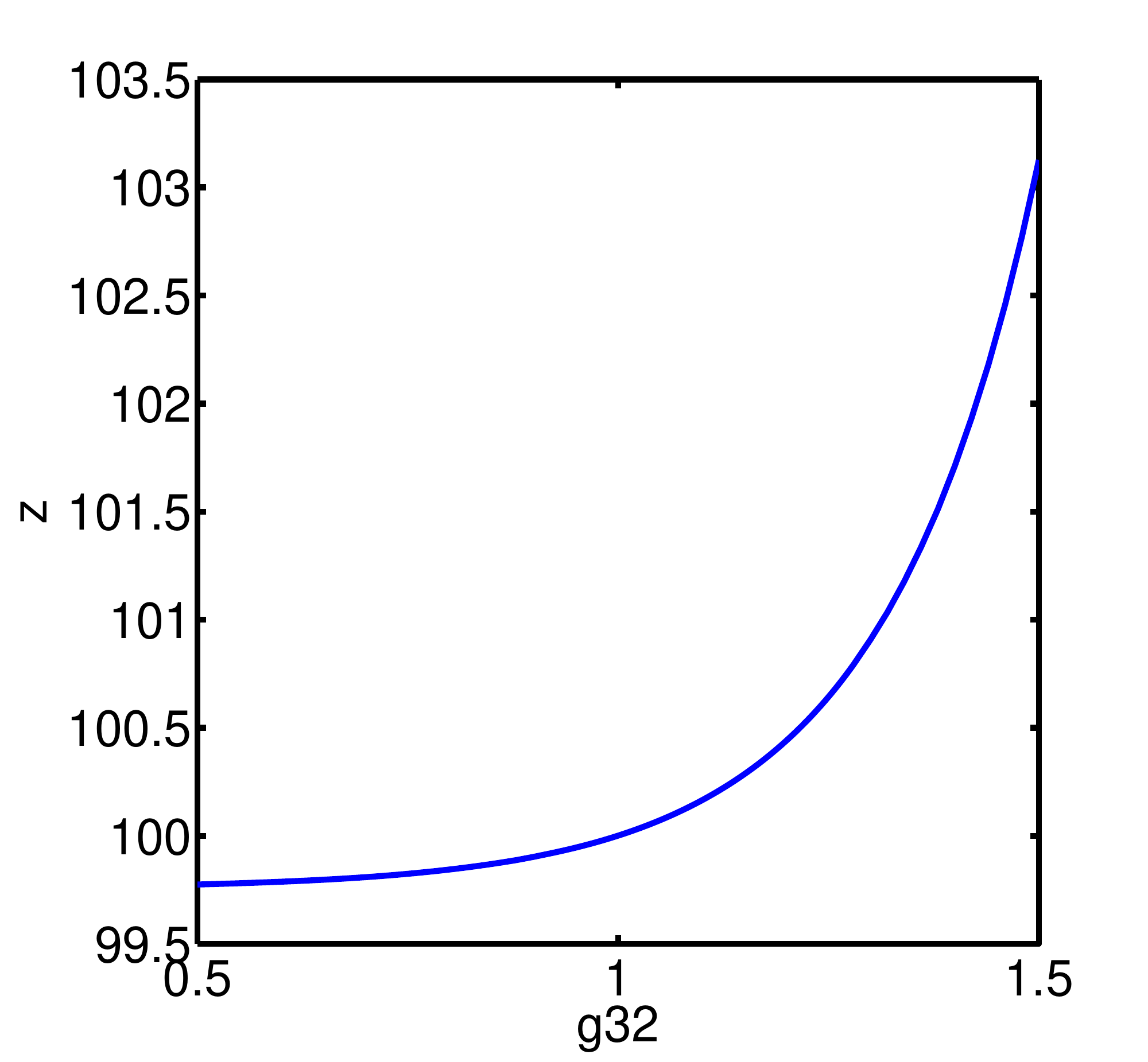}\\
  {{ \bf Figure 7.} The steady state bone volume, $z_{ss}$, computed as a function of the effectiveness of pre-osteoblast autocrine signaling for pre-osteoblast proliferation, $g_{32}$, with all other parameters held at baseline values.}%\label{figFive}
 \end{center}
\end{figure}

\begin{figure}[!ht]
  \begin{center}
  \includegraphics[width=5in]{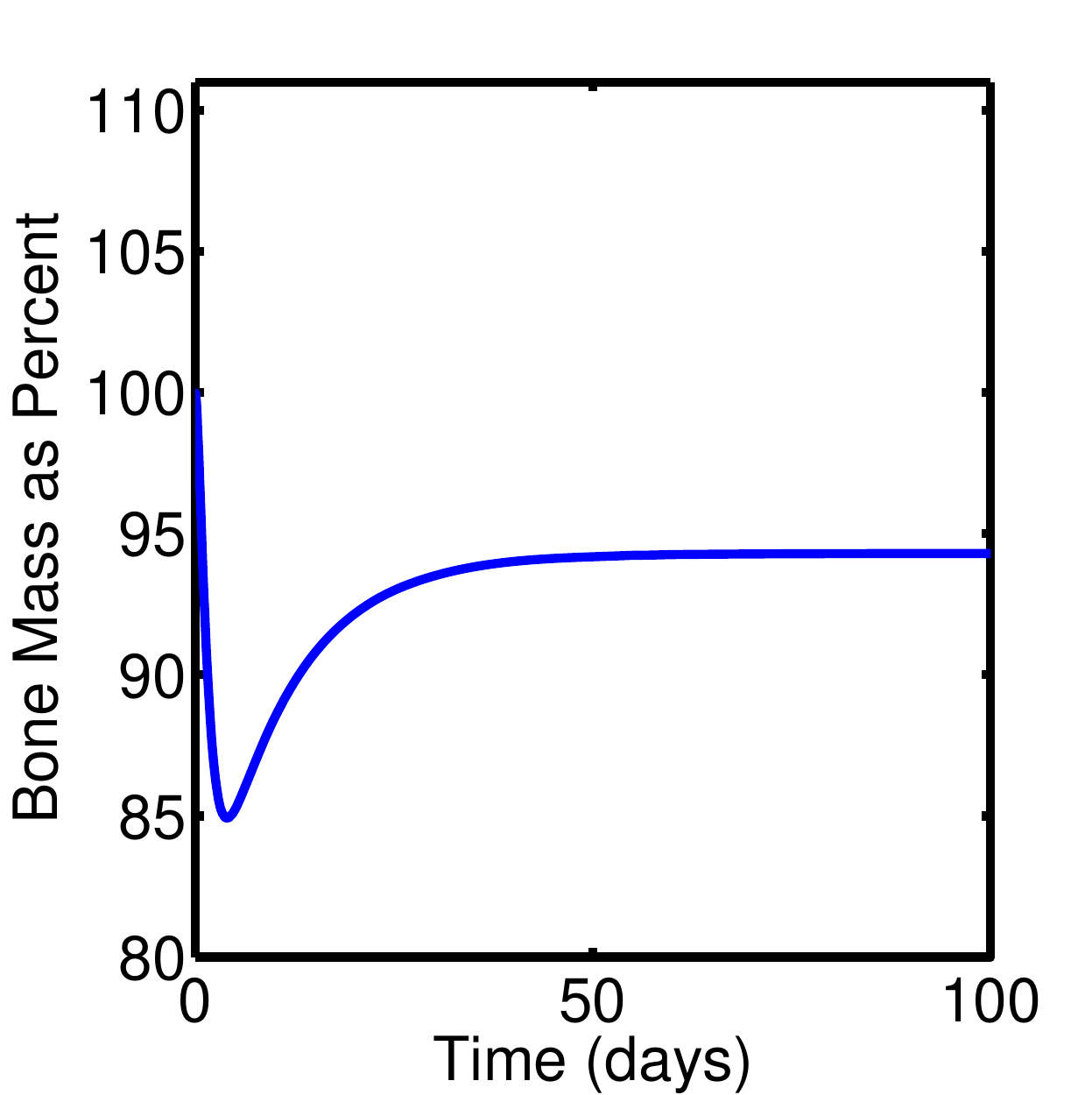}\\
  {{ \bf Figure 8.}  We simulate the loss of bone mass associated with over resorption in conjunction with a bone degenerative disease, modeled by a slight increase in the value of $\alpha_{4}$ with all other parameters set to the baseline values listed in table 2. }%\label{figFive}
 \end{center}
\end{figure}

\begin{figure}[!ht]
  \begin{center}
  \includegraphics[width=5in]{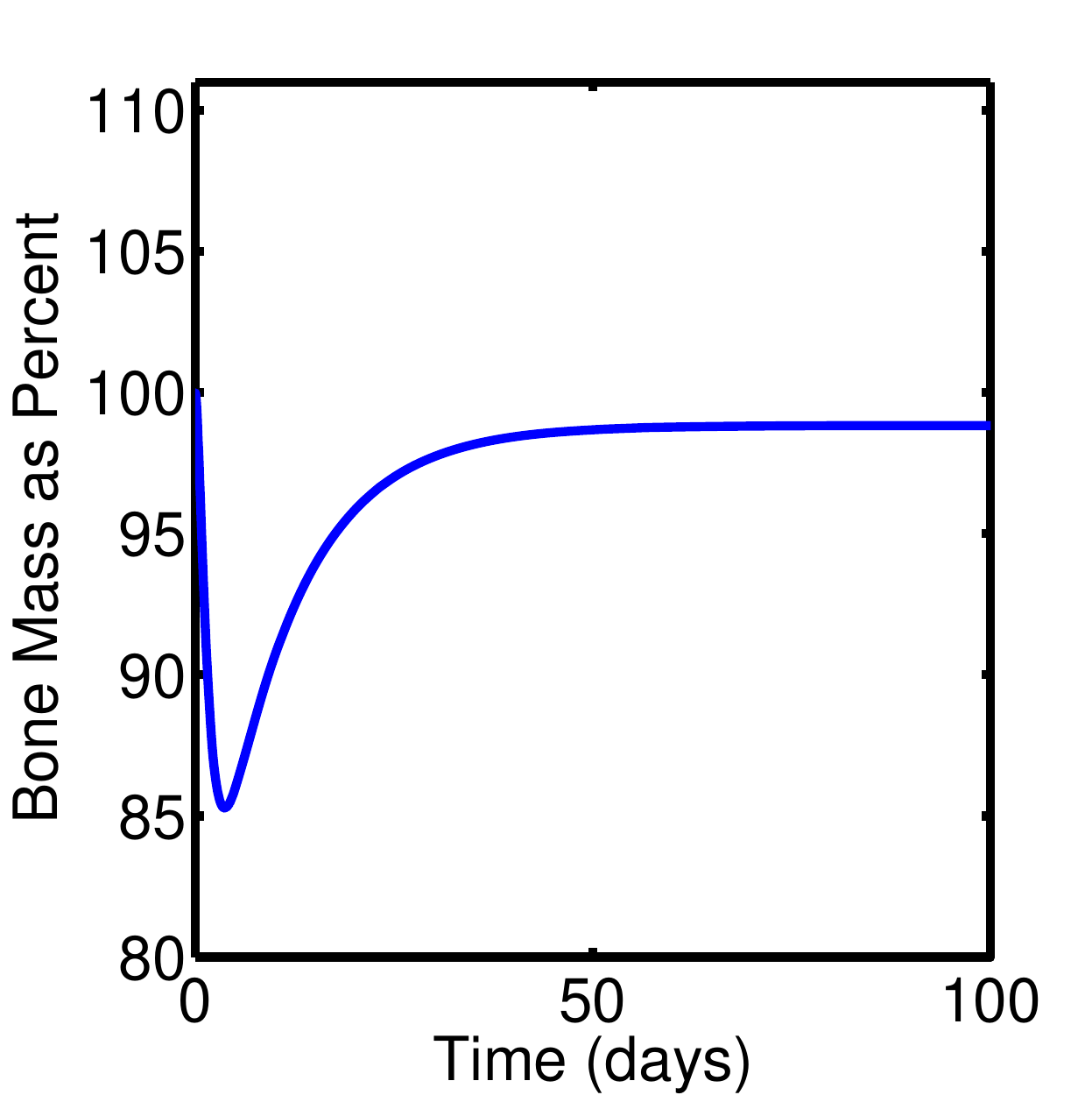}\\
  {{ \bf Figure 9.} This figure shows simulation results of treating pathological bone remodeling, as simulated in figure 5, via the addition of an anti-sclerostin drug. This results in a dose-dependent increase in bone mass. Treatment with an anti-sclerostin drug is modeled by modifying the appropriate signaling mechanisms, that is, by modifying the appropriate exponents $g_{22}$ and $g_{44}$ in the power law approximation.  }%\label{figFive}
\end{center}
\end{figure}

\section*{Tables}
%\begin{table}[!ht]
%\caption{
%\bf{Table title}}
%\begin{tabular}{|c|c|c|}
%table information
%\end{tabular}
%\begin{flushleft}Table caption
%\end{flushleft}
%\label{tab:label}
% \end{table}

\begin{table}[!ht]
  \caption{\bf{Definitions of Symbols Used in the Paper}}
  \label{symbols}
  \centering
  \begin{tabular}{|l|l|}
  \hline
  Symbol & Definition \\
  \hline
  $S$ & Number of osteocytes at a given time $t$\\
  $P$ & Number of pre-osteoblasts at a given time $t$\\
  $B$ & Number of osteoblasts at a given time $t$\\
  $C$ & Number of osteoclasts at a given time $t$\\
  $z$ & Bone volume at a given time $t$\\
  $\ao$ & Osteoblast embedding rate\\
  $\at$ & Differentiation rate of pre-osteoblast precursors\\
  $\att$ & Pre-osteoblast proliferation rate\\
  $\bo$ & Differentiation rate of pre-osteoblasts\\
  $\delta$ & Apoptosis of pre-osteoblasts\\
  $\bt$ & Osteoblast apoptosis\\
  $\af$ & Differentiation rate of osteoclast precursors\\
  $K_{S}$ & Critical value of osteocyte population\\
  $k_{1}$ & Bone resorption rate\\
  $k_{2}$ & Bone formation rate\\
  $g_{31}$ & Effectiveness of osteoblast autocrine signaling\\
  $g_{21}$ & Effectiveness of osteocyte paracrine signaling of pre-osteoblasts\\
  $g_{22}$ & Effectiveness of sclerostin regulation of osteoblastogenesis \\
  $g_{32}$ & Effectiveness of pre-osteoblast autocrine signaling\\
  $g_{41}$ & Effectiveness of osteocyte paracrine signaling of osteoclasts\\
  $g_{42}$ & Effectiveness of pre-osteoblast paracrine signaling of osteoclasts\\
  $g_{43}$ & Effectiveness of osteoblast paracrine signaling of osteoclasts\\
  $g_{44}$ & Effectiveness of sclerostin regulation of osteoclastogenesis \\
  $f_{12}$ & Effectiveness of pre-osteoblast paracrine signaling of osteoblasts\\
  $f_{14}$ & Effectiveness of osteoclast paracrine signaling of osteoblasts\\
  $f_{23}$ & Effectiveness of osteoblast autocrine signaling for apoptosis\\
  $f_{34}$ & Effectiveness of osteoclast autocrine signaling for apoptosis\\
  $\theta(\cdot)$ & Heaviside function \\
  \hline
  \end{tabular}
\end{table}

\begin{table}[!ht]
  \caption{\bf{Parameter Values}}
  \label{parameters}
  \centering
  \begin{tabular}{|p{2.0in}|p{1.2in}|p{1.2in}|} \hline
Parameter & Value & Units \\ \hline
$\alpha $${}_{1}$ & 0.5 & per day \\ \hline
$\alpha $${}_{2}$ & 0.1 & per day \\ \hline
$\alpha $${}_{3}$ & 0.1 & per day \\ \hline
$\beta $${}_{1}$ & 0.1 & per day \\ \hline
$\delta $ & 0.1 & per day \\ \hline
$\beta $${}_{2}$ & 0.1 & per day \\ \hline
$\alpha $${}_{4}$ & 0.1 & per day \\ \hline
K${}_{S}$ & 200 & number of cells \\ \hline
k${}_{1}$ & 0.7 & \% volume per day \\ \hline
k${}_{2}$ & 0.015445 & \% volume per day \\ \hline
g${}_{31}$ & 1 & dimensionless \\ \hline
g${}_{21}$ & 2 & dimensionless \\ \hline
g${}_{22}$ & 1 & dimensionless \\ \hline
g${}_{32}$ & 1 & dimensionless \\ \hline
g${}_{41}$ & 1 & dimensionless \\ \hline
g${}_{42}$ & 1 & dimensionless \\ \hline
g${}_{43}$ & -1 & dimensionless \\ \hline
g${}_{44}$ & 1 & dimensionless \\ \hline
f${}_{12}$ & 1 & dimensionless \\ \hline
f${}_{14}$ & 1 & dimensionless \\ \hline
f${}_{23}$ & 1 & dimensionless \\ \hline
f${}_{34}$ & 1 & dimensionless \\ \hline
$\varepsilon $ & 1 & number cells \\ \hline
\end{tabular}
\end{table}

\begin{table}[!ht]
  \caption{{\bf Effectiveness of RANKL Expression.} }
  \label{rankl}
  \centering
  \begin{tabular}{|p{2.1in}|p{0.3in}|} \hline
\textit{g}${}_{41}$ & \textit{g}${}_{42}$\newline  \\ \hline
1\newline 0.22\newline 0.24\newline 0.26\newline 0.28\newline 0.3\newline 0.6\newline 0.62\newline 0.82\newline 0.84\newline 1.14\newline 1.24\newline 1.34\newline 1.42\newline 1.66 & 1\newline 1.84\newline 1.82\newline 1.8\newline 1.78\newline 1.76\newline 1.44\newline 1.42\newline 1.2\newline 1.18\newline 0.84\newline 0.72\newline 0.6\newline 0.5\newline 0.18 \\ \hline
\end{tabular}
\begin{flushleft}Values of the effectiveness of RANKL expression by osteocytes ($g_{41}$), and pre-osteoblasts ($g_{42}$) corresponding to normal, targeted, bone remodeling.
\end{flushleft}
\end{table}

\end{document}